\documentclass[10pt,conference]{IEEEtran}

\usepackage{color,graphicx,amsmath,amssymb,epsfig,mathrsfs,cite,multirow,array}
\usepackage[caption=false,font=footnotesize]{subfig}
\usepackage{algorithm2e, setspace}
\usepackage{enumitem}

\everymath{\textstyle}
\let\displaystyle\textstyle
\usepackage[flushleft]{threeparttable}

\newcommand{\squeezeup}{\vspace{-0.5cm}}
\newcommand{\MI}{\mathrm{MI}}
\newcommand{\I}{\mathrm{I}}

\IEEEoverridecommandlockouts

\begin{document}
\title{Data-Driven Estimation Of Mutual Information Between Dependent Data}
\author{\IEEEauthorblockN{Rakesh Malladi, Don H Johnson, and Behnaam Aazhang}
\IEEEauthorblockA{Department of Electrical and Computer Engineering, Rice University, Houston, USA}
\thanks{This work is funded in part by grant 1406447 from National Science Foundation.  The authors can be reached at 
\{Rakesh.Malladi, dhj, aaz\}@rice.edu}}

\maketitle

\begin{abstract}
``To be considered for the 2017 IEEE Jack Keil Wolf ISIT Student Paper Award." We consider the problem of estimating mutual information between dependent data, an important problem in many science and engineering applications. We  propose a data-driven, non-parametric estimator of mutual information in this paper. The main novelty of our solution lies in transforming the data to frequency domain to make the problem tractable. We define a novel metric--mutual information in frequency--to detect and quantify the dependence between two random processes across frequency using Cram\'{e}r's spectral representation. Our solution calculates mutual information as a function of frequency to estimate the mutual information between the dependent data over time. We validate its performance on linear and nonlinear models. In addition, mutual information in frequency estimated as a part of our solution can also be used to infer cross-frequency coupling in the data. 

\end{abstract}

\section{Introduction}
Identifying the dependence relationships among multiple data streams sampled from a system is a problem of interest in many science and engineering applications. Typically, we sample data for a finite duration and are interested in detecting and quantifying the dependence between the data. For instance, given electrocorticographic (ECoG) recordings from different spatial locations in the brain, we are interested in detecting if the activity at two different locations is independent or not, and if not, quantifying their dependence \cite{friston1994}. Mutual information (MI), introduced by Shannon in 1948 \cite{shannon1948}, is a powerful and well developed tool that has been used to detect if two data streams are independent and to quantify any dependence using a non-negative scalar \cite{cover2012}. In this paper, we focus on estimating mutual information between two data steams from a finite number of samples.


Estimating mutual information from independent and identically distributed (i.i.d.) data is a well-studied problem and a good review of the popular algorithms is provided in \cite{wang2009}. However, data samples recorded from real-world systems like brain are usually dependent across time. Even with dependent data, if the underlying model is known to be linear and Gaussian, mutual information can be estimated using the power spectral density \cite{cover2012} or coherence \cite{pinsker1960}.  In most real-world problems, the underlying model is not known. The main contribution of this paper is that we develop a novel data-driven algorithm to estimate the MI between two stochastic processes from dependent data without imposing any parametric model assumptions.


The key idea behind our approach is to estimate the MI by transforming the time-domain random processes to stochastic processes in frequency using Cram\'{e}r's spectral representation \cite{pinsker1960, larson1979, cramer2013}. We then define a novel measure of dependency in frequency called `mutual information in frequency' between different frequency components of stochastic processes, which is equivalent to coherence for Gaussian processes. Mutual information in frequency can be viewed as a generalization of coherence to non-Gaussian processes. Note that we developed mutual information in frequency to identify cross-frequency coupling (dependence in data across frequency) in neuroscience and applied to ECoG recordings from brains of epilepsy patients \cite{malladi2017}. We estimate MI variations between the two observed data streams in frequency and in time by dividing the data into sufficiently long non-overlapping windows. The MI between the two processes is calculated by estimating the MI between the groups of frequencies in the two processes with statistically significant MI in frequency. The  proposed MI estimator converges to the true value for Gaussian data and our simulation results demonstrate it works well for nonlinear models. In addition to quantifying the dependence between the data streams, MI in frequency, calculated as a step in our proposed algorithm,  identifies the frequency bands containing the common information between the underlying random processes. Identifying  cross-frequency coupling plays an important role in understanding neuronal computation and learning \cite{aru2015}. 

\section{Problem Statement} 
Let $X$ and $Y$ denote two discrete-time random processes. We consider the problem of estimating the MI between them, $\hat{\I}\left(X;Y\right)$, from dependent, but identically distributed data points, $\left(x[n], y[n]\right) \in \mathbb{R}^2$, $n=0,1,\cdots,\left(N-1\right)$.
The mutual information rate between two stochastic processes, $X$ and $Y$ is defined as 
\begin{align} \label{eq:mi_rate}
\I\left(X;Y\right) =  \lim\limits_{N \rightarrow \infty} \frac{1}{N}\I\left(X^{N};Y^{N}\right),
\end{align}
where $X^{N} \!\!=\!\! \left(\!X[0],\!\cdots\!,\! X[N-1] \right),\! Y^{N} \!\!=\!\! \left(\!Y[0], \!\cdots\!, \!Y[N-1] \right)$ are $N$-element random vectors. We need to estimate the mutual information between two $N$-element random vectors to estimate $\hat{\I}\left(X;Y\right)$ using \eqref{eq:mi_rate}. This problem can be solved if the underlying model is linear and Gaussian by estimating MI in the frequency domain \cite{pinsker1960, cover2012}. However, explicit calculations for most real-world applications is impossible, since the underlying model is unknown and can potentially be highly nonlinear \cite{wang2009}. The main idea behind our solution is that computing mutual information in the frequency domain in a data-driven manner not only makes the problem much more computationally tractable, but also provides insights into the nature of dependence.

To achieve this, we first define mutual information in frequency, ${\MI}_{XY}\left(\lambda_i, \lambda_j\right)$, a novel metric that detects and quantifies the statistical dependence between $\lambda_i$ frequency component of $X$ and $\lambda_j$ of $Y$ using Cram\'{e}r's spectral representation \cite{pinsker1960, larson1979, cramer2013}, where $\lambda_i, \lambda_j$ are the normalized frequencies. We then propose a data-driven estimator, based on k-nearest neighbors (k-NN), to estimate MI in frequency from data. The statistical significance of the resulting estimates is tested by permuting the data under null hypothesis. With $\Lambda_x$ and $\Lambda_y$ denoting the frequency components having statistically significant non-zero values of mutual information, our proposed estimate of mutual information between $X$ and $Y$, $\hat{\I}\left(X;Y \right)$, is estimated using the information between all the frequencies in $\Lambda_x$ and $\Lambda_y$ of $X$ and $Y$ respectively. 

\section{Mutual Information in Frequency}
\label{sec:MI_def}
Consider two discrete-time real-valued stochastic processes $X$ and $Y$. We proceed to define the mutual information in frequency, $\MI_{XY}\left(\lambda_i,\lambda_j\right)$, between $\lambda_i^{th}$ and $\lambda_j^{th}$ component of $X$ and $Y$ respectively, where $\lambda_i, \lambda_j \in \left[0, 1\right]$ are the normalized frequencies. Assuming $X$ is a second order stationary, mean-square continuous, zero mean process, there exists an orthogonal increment process $\widetilde{X}\left(\lambda_i\right)$, called the spectral process or the Cram\'{e}r's representation \cite{pinsker1960, larson1979, cramer2013} of $X$ at $\lambda_i$, that satisfies
\begin{align}\label{eq:cramer_rep}
\!\!X\left[n\right] \!\!=\!\! \int\limits_{0}^1\! e ^{j2\pi\lambda_i n}\! d\widetilde{X}\left(\lambda_i\right), \: \text{and} \: \mathbb{E}\big[|d\widetilde{X}\left(\lambda_i\right)|^2 \big] \!\! = \!\! dS_X\left(\lambda_i\right)\!,
\end{align}
where $S_X\left(\lambda_i\right)$ is the spectral distribution function of $X$ and $d\widetilde{X}\left(\lambda_i\right) \in \mathbb{C}$ is called the spectral process increments of $X$ at normalized frequency $\lambda_i$. Similarly let $d\widetilde{Y}\left(\lambda_j\right) \in \mathbb{C}$ denote the increments of spectral processes of $Y$ at normalized frequencies $\lambda_j$. Let $\mathrm{P}\big( d\widetilde{X}_R\left( \lambda_i\right),d\widetilde{X}_I\left( \lambda_i\right) ,d\widetilde{Y}_R\left( \lambda_j\right),d\widetilde{Y}_I\left( \lambda_j\right) \big)$ be the joint probability density of the four dimensional random vector of the real and imaginary parts of $d\widetilde{X}\left(\lambda_i\right)$ and $d\widetilde{Y}\left(\lambda_j\right)$. Also, let $\mathrm{P}\big(d\widetilde{X}_R\left( \lambda_i\right),d\widetilde{X}_I\left( \lambda_i\right)\big)$ and $\mathrm{P}\big(d\widetilde{Y}_R\left( \lambda_j\right),d\widetilde{Y}_I\left( \lambda_j\right)\big)$ denote the corresponding two-dimensional marginal densities. The mutual information between $X$ at frequency $\lambda_i$ and $Y$ at $\lambda_j$ is defined as \begin{align} \label{MI_freq_def}
& \MI_{XY}\left(\lambda_i,\lambda_j\right) \nonumber \\
& = \I\big(\big\{d\widetilde{X}_R\left( \lambda_i\right),d\widetilde{X}_I\left( \lambda_i\right) \big\}; \big\{d\widetilde{Y}_R\left( \lambda_j\right),d\widetilde{Y}_I\left( \lambda_j\right) \big\} \big),
\end{align} 
where $\I\left(\left\{\cdot,\cdot\right\};\left\{\cdot,\cdot\right\} \right)$ is the standard mutual information between two pairs of two dimensional  real-valued random vectors \cite{cover2012}. 
The MI between two different frequencies $\lambda_i$, $\lambda_j$ in the same process $Y$ is similarly defined as
\begin{align}\label{MI_freq_def_one_proc}
& \MI_{YY}\left(\lambda_i,\lambda_j\right) \nonumber \\  
& = \I\big(\big\{d\widetilde{Y}_R\left( \lambda_i\right),d\widetilde{Y}_I\left( \lambda_i\right) \big\}; \big\{d\widetilde{Y}_R\left( \lambda_j\right),d\widetilde{Y}_I\left( \lambda_j\right) \big\} \big).
\end{align}
MI in frequency defined in \eqref{MI_freq_def}, \eqref{MI_freq_def_one_proc} is a non-negative number that is zero if the two frequency components are independent, and, if they are dependent, quantifies the common information between them. MI in frequency between two processes \eqref{MI_freq_def} is not symmetric in general: $\MI_{XY}\left(\lambda_i,\lambda_j\right) \neq \MI_{XY}\left(\lambda_j,\lambda_i\right)$. However, it is symmetric within a process: $\MI_{YY}\left(\lambda_i,\lambda_j\right) = \MI_{YY}\left(\lambda_j,\lambda_i\right)$.  The MI between the components of $Y$ at frequencies $\lambda_j$ and $\lambda_j$, $\MI_{YY}\left(\lambda_j,\lambda_j\right)$, is $\infty$, a consequence of the fact that $\big[d\widetilde{Y}_R\left( \lambda_j\right),d\widetilde{Y}_I\left( \lambda_j\right) \big]$ is a continuous-valued random vector. Mutual information in frequency is equivalent to coherence for linear, Gaussian models and can be viewed as a generalization of coherence for non-Gaussian processes. More details about the mutual information in frequency metric and its application in neuroscience are given in \cite{malladi2017}.

\section{Data-Driven MI Estimator} \label{sec:mi_est}
The proposed data-driven MI estimation algorithm takes in $N$ samples of $X$ and $Y$ as input and outputs the mutual information between $X$ and $Y$, $\hat{\I} \left(X; Y \right)$, without assuming a parametric model for the relationship between data.

\RestyleAlgo{boxruled}
\begin{algorithm}[ht]
\setstretch{1.15}
\SetAlgoLined
\caption{Mutual Information Estimator\vspace{.05cm}}\label{mi_est_algo}
\textit{Data} - $\left(x\left[n\right], y\left[n\right] \right)$, for $x\left[n\right], y\left[n\right] \in \mathbb{R}, n\in[0,N-1].$ \\
\textit{Output} - $\hat{\I} \left(X; Y \right)$ \\
\textit{Algorithm}
\begin{enumerate}[leftmargin=*]
\item[\textit{A)}] Select an appropriate value for $N_f$ and divide the data into $N_s$ windows such that $N_f\times N_s = N$.
\item[\textit{B)}] Estimate  $\widehat{\MI}_{XY}\left(\lambda_i, \lambda_j \right)$, 
where $\lambda_i = \frac{i}{N_f}, \lambda_j=\frac{j}{N_f}$, $\forall \left(i, j\right) \ni i, j \in \left[0, N_f-1 \right]$.
\item[\textit{C)}] Find the sets $\Lambda_x, \Lambda_y$,  such that $\widehat{\MI}_{XY}\left(\lambda_{i_p}, \lambda_{j_q} \right)$ is statistically significant $\forall \lambda_{i_p} \in \Lambda_x, \lambda_{j_q} \in \Lambda_y$, where $i_p, j_q \in \left[0, N_f-1 \right]$. Let $P, Q$ respectively denote the cardinality of $\Lambda_x, \Lambda_y$.
  \item[\textit{D)}] Let $d\widetilde{X}\left(\Lambda_x \right) = \big[d\widetilde{X}\left(\lambda_{j_1} \right), \cdots, d\widetilde{X}\left(\lambda_{j_P} \right) \big] \in \mathbb{R}^{2P}$ and $d\widetilde{Y}\left(\Lambda_y \right) = \big[d\widetilde{Y}\left(\lambda_{l_1} \right), \cdots, d\widetilde{Y}\left(\lambda_{l_Q} \right) \big] \in \mathbb{R}^{2Q}$. The mutual information between $X$ and $Y$ is given by
  \begin{align}
  \hat{\I} \left(X; Y \right) = \frac{1}{\max(P,Q)} \hat{\I} \left(d\widetilde{X}\left(\Lambda_x \right);  d\widetilde{Y}\left(\Lambda_y \right)\right), \nonumber
  \end{align}
  where the MI on right hand side is estimated from $N_s$ i.i.d samples using any nonparametric MI estimator \cite{wang2009}.
  \end{enumerate}
\end{algorithm}
\subsection{Choosing $N_f$}
The first step of the algorithm is finding the appropriate value for $N_f$, which essentially encodes the length of dependence in the data and we assume data in different windows are independent of each other. Ideally, consecutive windows should be separated to ensure no dependence across windows, but our simulation results demonstrate that no separation between windows doesn't affect performance significantly. In addition, $N_f$ also determines the frequency resolution of our MI in frequency estimates. Assuming the underlying distribution is stationary and satisfies a mixing assumption  \cite{brillinger2001}, the $N$ samples of $X$ are split into $N_s$ non-overlapping windows with $N_f = \frac{N}{N_s}$ data points in each window. Let us denote the samples in $l^{th}$ window of $X$ and $Y$ respectively by two $N_f$ element one-dimensional vectors, $\mathbf{x}^{l}$ and $\mathbf{y}^{l}$, for $l=1,2,\cdots,N_s$.

\subsection{Data-Driven Estimator of MI in Frequency} \label{sec:ksg_est}
The second step of the algorithm involves estimating mutual information in frequency,  $\widehat{\MI}_{XY}\left(\lambda_i, \lambda_j \right)$, between $\lambda_i$ component of $X$ and $\lambda_j$ component of $Y$, where $\lambda_i = \frac{i}{N_f}, \lambda_j=\frac{j}{N_f}$, $\forall \left(i, j\right) \ni i, j \in \left[0, N_f-1 \right]$. The data-driven MI in frequency estimator, $\widehat{\MI}_{XY}\left(\lambda_i, \lambda_j \right)$, consists of two steps: first estimating samples of spectral process increments, $d\widetilde{X}\left(\lambda_i\right)$ and $d\widetilde{Y}\left(\lambda_j\right)$ at $\lambda_i$ and $\lambda_j$ respectively and then estimating MI from these samples using a data-driven estimator.

\subsubsection{Estimation of Samples of Spectral Process Increments} \label{sec:est_samples}
Let us focus on estimating samples of the random variable $d\widetilde{X}\left(\lambda_i\right)$. Let $\mathcal{F}\left\{\mathbf{x}^l\right\}\left(\alpha \right)$ denote the discrete-time Fourier transform (DTFT) of $\mathbf{x}^{l}$ at normalized frequency $\alpha$. For $\lambda_i = \frac{i}{N_f} \in \left[0,1\right] \text{and} \: i \in \left[0, N_f-1 \right]$, let us define $d\widetilde{x}^{l}\left(\lambda_i\right)$ and integrated Fourier spectrum, $\widetilde{x}^{l}\left(\lambda_i\right)$, by
\begin{align} \label{spec_proc_inc_est}
\!\!d\widetilde{x}^{l}\left(\lambda_i\right) = \mathcal{F}\left\{\mathbf{x}^l \right\}\left(\lambda_i \right) \:  \text{and} \: \widetilde{x}^{l}\left(\lambda_i\right) = \sum\limits_{m=0}^{i} \mathcal{F}\left\{\mathbf{x}^l \right\}\left(\lambda_m \right).
\end{align}
It is shown in \cite{brillinger2001} that the random variable for which $\widetilde{x}^{l}\left(\lambda_i\right)$ is just one realization tends to the spectral process of $X$ at $\lambda_i$ in mean of order $\gamma$, for any $\gamma > 0$, as the number of samples goes to infinity. Also, $d\widetilde{x}^{l}\left(\lambda_i\right)$, which is the increment in $\widetilde{x}^l\left(\lambda_i\right)$ between $\lambda_i$ and $\lambda_i+d\lambda$, is just the DTFT of the samples in window $i$. 
Calculating the DTFT with the FFT for each of the $N_s$ windows separately yields an $N_f \times N_s$ matrix, whose $i^\mathrm{th}$ row, $\mathbf{d\widetilde{x}}\left(\lambda_i\right) = \left[d\widetilde{x}^{1}\left(\lambda_i\right),d\widetilde{x}^{2}\left(\lambda_i\right),\cdots,d\widetilde{x}^{N_s}\left(\lambda_i\right)\right]$ is the complex-valued vector containing $N_s$ samples of  $d\widetilde{X}\left(\lambda_i\right)$, the spectral process increments of $X$ at $\lambda_i=\frac{i}{N_f}$. The $l^{th}$ element of $\mathbf{d\widetilde{x}}\left(\lambda_i\right)$, $d\widetilde{x}^{l}\left(\lambda_i\right) = d\widetilde{x}_R^{l}\left(\lambda_i\right) + i d\widetilde{x}_I^{l}\left(\lambda_i\right)$,  is a particular realization of $d\widetilde{X}\left(\lambda_i\right)$. A similar procedure is used to obtain the $N_s$ samples of the spectral process increments of $Y$ at $\lambda_j= \frac{j}{N_f},  j \in \left[0, N_f-1 \right]$ and the resulting data samples are denoted by $\mathbf{d\widetilde{y}}\left(\lambda_j\right) = \left[d\widetilde{y}^{1}\left(\lambda_j\right),d\widetilde{y}^{2}\left(\lambda_j\right),\cdots,d\widetilde{y}^{N_s}\left(\lambda_j\right)\right]$.

\subsubsection{Data-Driven MI in Frequency Estimator}
${\MI}_{XY}\left(\lambda_i, \lambda_j \right)$ is now estimated from $d\widetilde{x}^l\left(\lambda_i \right)\in \mathbb{R}^2$ and $d\widetilde{y}^l\left(\lambda_j \right)\in \mathbb{R}^2$, for $l=1, 2, \cdots, N_s$. A good review of various non-parametric estimators of mutual information is provided in \cite{wang2009}. We compared the performance of a plug-in kernel density estimator (KDE) \cite{scott2015, wang2009} and a k-nearest neighbor based estimator (k-NN) \cite{kraskov2004, wang2009} for mutual information. We found that the k-NN based estimator outperforms the KDE based estimator in terms of bias and rate of convergence \cite{malladi2017}. 
We apply the first version of the algorithm in \cite{kraskov2004} to two-dimensional random variables $d\widetilde{X}\left(\lambda_i\right)$ and $d\widetilde{Y}\left(\lambda_j\right)$ to compute $\widehat{\MI}_{XY}\left(\lambda_i, \lambda_j \right)$. Consider the joint four dimensional space $\big(d\widetilde{X}\left(\lambda_i\right), d\widetilde{Y}\left(\lambda_j\right)\big) \in \mathbb{R}^4$. The distance between two data points with indices $l_1, l_2 \in \left[1, N_s \right]$ is calculated using the infinity norm, according to $\max \left\{ \|d\widetilde{x}^{l_1}\left(\lambda_i\right) - d\widetilde{x}^{l_2}\left(\lambda_i\right) \|, \|d\widetilde{y}^{l_1}\left(\lambda_j\right) - d\widetilde{y}^{l_2}\left(\lambda_j\right) \| \right\}$. Let $\epsilon_l$ denote the distance between the data sample $\big(d\widetilde{x}^{l}\left(\lambda_i\right), d\widetilde{y}^{i}\left(\lambda_j\right)\big)$ and its $K^{th}$ nearest neighbor, for $l =1, 2, \cdots, N_s$. We used $K=3$ in this paper. Let $n_x^l$ and $n_y^l$ denote the number of samples of $d\widetilde{X}\left(\lambda_i\right)$ and $d\widetilde{Y}\left(\lambda_j\right)$ within an infinity norm ball of radius less than $\epsilon_l$ centered at $d\widetilde{x}^{l}\left(\lambda_i\right)$ and $d\widetilde{y}^{i}\left(\lambda_j\right)$ respectively. The mutual information in frequency between $X$ and $Y$ at normalized frequencies $\lambda_i$ and $\lambda_j$ is given by
\begin{align} \label{eq_knn_est}
\widehat{\MI}_{XY}\left(\lambda_i,\lambda_j\right) & = \psi\left(K\right) + \psi\left(N_s\right) \nonumber \\
& - \frac{1}{N_s}\sum\limits_{l=1}^{N_s} \left(\psi\left(n_x^l + 1\right) + \psi\left(n_y^l + 1 \right) \right),
\end{align}
where $\psi\left(\cdot\right)$ is the Digamma function. 

\subsubsection{Statistical Significance Testing}
To test the statistical significance of this estimate, we permute the elements in the $N_s$ samples of $d\widetilde{X}\left(\lambda_i\right)$ randomly and estimate the MI in frequency between the permuted vector and the $N_s$ samples of $d\widetilde{Y}\left(\lambda_j\right)$ using \eqref{eq_knn_est}. We permute $N_p$ times to obtain $N_p$ permuted MI in frequency estimates, under the null hypothesis of independence. The permuted MI estimates will be almost zero, since the permutations make the spectral processes almost independent. If the actual MI estimate, $\widehat{\MI}_{XY}\left(\lambda_i,\lambda_j\right)$, is judged larger than all the permuted $N_p$ estimates, then there is a statistically significant dependence between the processes at these two frequencies.

\subsection{Identifying Coupled Frequencies}
The third step in the proposed algorithm involves identifying the set of frequency components in $X$ and in $Y$, denoted by $\Lambda_x$ and $\Lambda_y$ respectively, that have statistically significant MI in frequency estimates. This can be graphically visualized by plotting the statistically significant MI in frequency estimates on a two-dimensional image grid (see Fig.~\ref{Fig:cosine_square}), whose rows and columns correspond to frequencies of $X$ and $Y$ respectively and identifying the frequency pairs  with significant MI in frequency estimates. This is a big positive feature of our solution. In addition to quantifying the dependence between $X$ and $Y$ by a non-negative scalar, we can also characterize the cross-frequency coupling between the data streams. There is a lot of interest in inferring cross-frequency coupling from data in areas like neuroscience \cite{aru2015} and our proposed MI estimator infers it along the way for free.

\subsection{Estimating Mutual Information}
The fourth and final step in the proposed algorithm estimates mutual information between the spectral process increments of $X$ and $Y$ at frequencies in $\Lambda_x$ and $\Lambda_y$ respectively. With $P, Q$ denoting the cardinality of $\Lambda_x, \Lambda_y$ respectively, let $d\widetilde{X}\left(\Lambda_x \right)$ and $d\widetilde{Y}\left(\Lambda_y \right)$ denote the $2P$ and $2Q$-dimensional random vector comprising the spectral process increments of $X$, $Y$ at all frequencies in $\Lambda_x$ and $\Lambda_y$ respectively. We already computed $N_s$ i.i.d. samples of these two random vectors to estimate pairwise MI in frequency estimates in step 2 of the algorithm. The desired MI estimate is computed from the mutual information between a $2P$ and $2Q$-dimensional random vector of spectral process increments, which is estimated using a k-nearest neighbor based estimator \cite{kraskov2004}, according to
  \begin{equation}\label{eq:mi_est}
  \hat{\I} \left(X; Y \right) = \frac{1}{\max\left(P,Q\right)} \hat{\I} \left(d\widetilde{X}\left(\Lambda_x \right);  d\widetilde{Y}\left(\Lambda_y \right)\right).
  \end{equation}

The MI estimator in \eqref{eq:mi_est} can be further simplified for linear, Gaussian models. Without loss of generality, consider two Gaussian processes $X$ and $Y$, related by 
\begin{align} \label{eq:linear_model}
y[n] = h[n]*x[n] + w[n],
\end{align} for some $h[n], w[n]$, where $h[n]$ is a linear time-invariant (LTI) filter and $W$ is colored Gaussian noise independent of $X$. For this class of models, \eqref{eq:mi_est} can be further simplified to
\begin{align}\label{eq:linear_model_mi_est}
\hat{\I}\left(X;Y\right) =  \frac{1}{N_f}\sum\limits_{i=0}^{N_f/2} \widehat{\MI}_{XY}\left(\lambda_i ; \lambda_i\right), \: \text{where} \: \lambda_i = \frac{i}{N_f}.
\end{align}
This result obtains because linear models do not introduce cross-frequency dependencies. Independently, we can also prove for this class of models that, MI between $X$ and $Y$ is related to MI in frequency according to \cite{malladi2017}
\begin{align}\label{eq:linear_model_mi}
\I\left(X;Y\right) = \int\limits_{0}^{0.5} \MI_{XY}\left(\lambda ; \lambda\right) d\lambda.
\end{align}
It is easy to see that the right hand side of \eqref{eq:mi_est} is just the Riemann sum of the integral on the right hand side of \eqref{eq:linear_model_mi}, which converges to the true value as $N_f$ tends to infinity. This implies the proposed estimator converges to the true value for Gaussian processes.

Note that the MI estimation algorithm proposed in this section does not make any parametric assumptions on the underlying model between $X$ and $Y$. The computation of MI via \eqref{eq:mi_est} can be greatly simplified by clustering the frequencies in $\Lambda_x$ and $\Lambda_y$ into groups without any significant dependencies across groups and using the chain rule of mutual information. In addition, if we observe after step 3 that significant MI in frequency estimates occur only at $\left(\lambda_i, \lambda_i \right),\! \forall i \!\in\! \big[0, N_f-1\big]$, then the MI can be estimated using \eqref{eq:linear_model_mi_est}.  

\section{Performance on Simulated Data}
We demonstrate the performance of the proposed estimator on data generated from four different models - two linear models and two nonlinear models. 

\subsection{Linear Models}
The data were generated from the model \eqref{eq:linear_model}, where $x[n], w[n]$ are i.i.d Gaussian noise with variances $\sigma_x^2, \sigma_w^2$ respectively and independent. We applied the proposed algorithm to estimate $\hat{\I}\left(X;Y\right)$ using $N_f=64$ and $N_s=10^4$ and averaged using $10$ different random number generator seeds for two different filters $h[n]$.
\subsubsection{Lowpass Filter} \label{sec:lowpass}
\begin{figure}[!t]
\centering
\subfloat[Two-tap lowpass filter]{
\includegraphics[width=0.45\columnwidth]{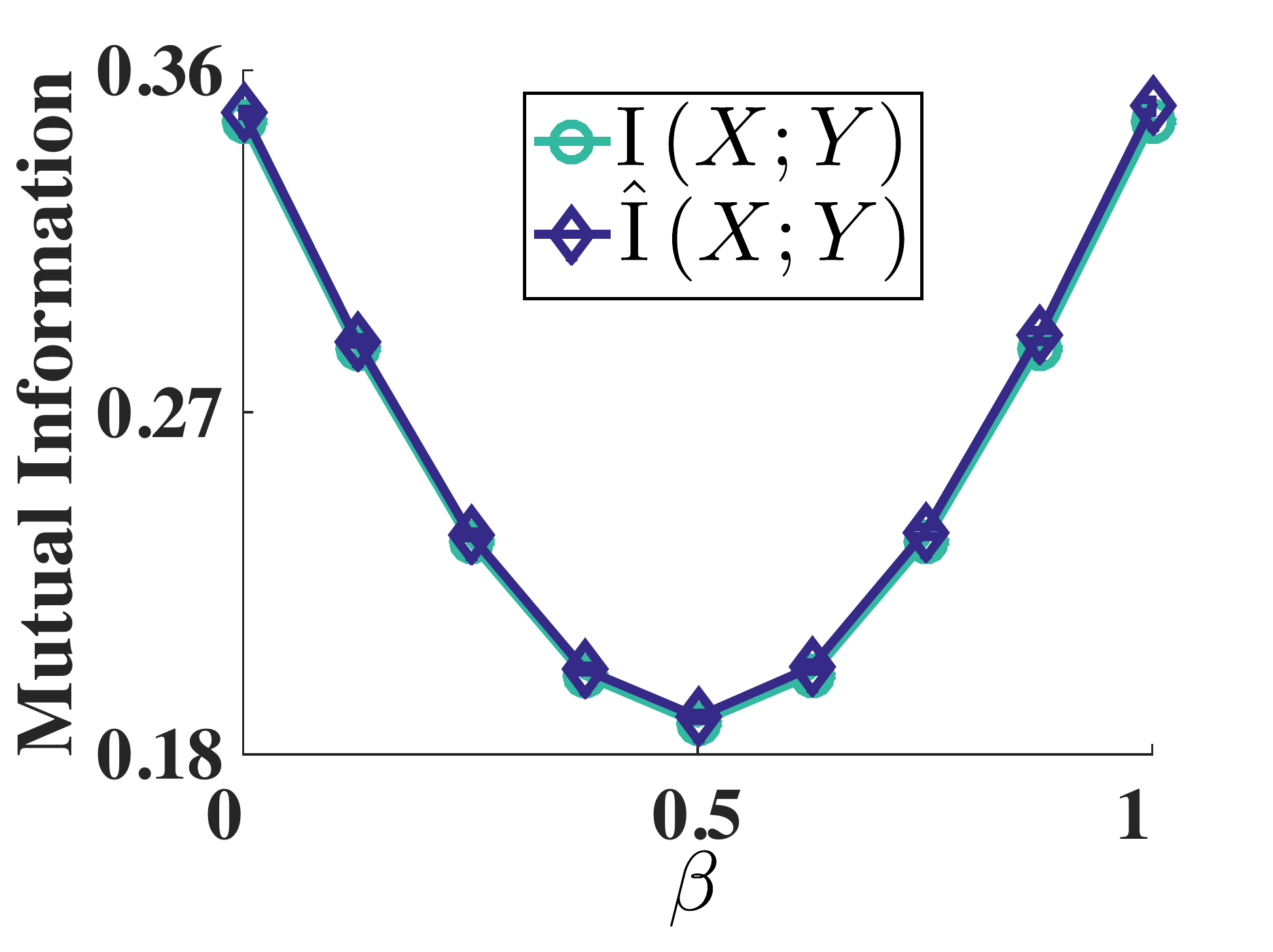}}
\hfill \vrule \hfill
\subfloat[33-tap bandpass filter]{
\includegraphics[width=0.45\columnwidth]{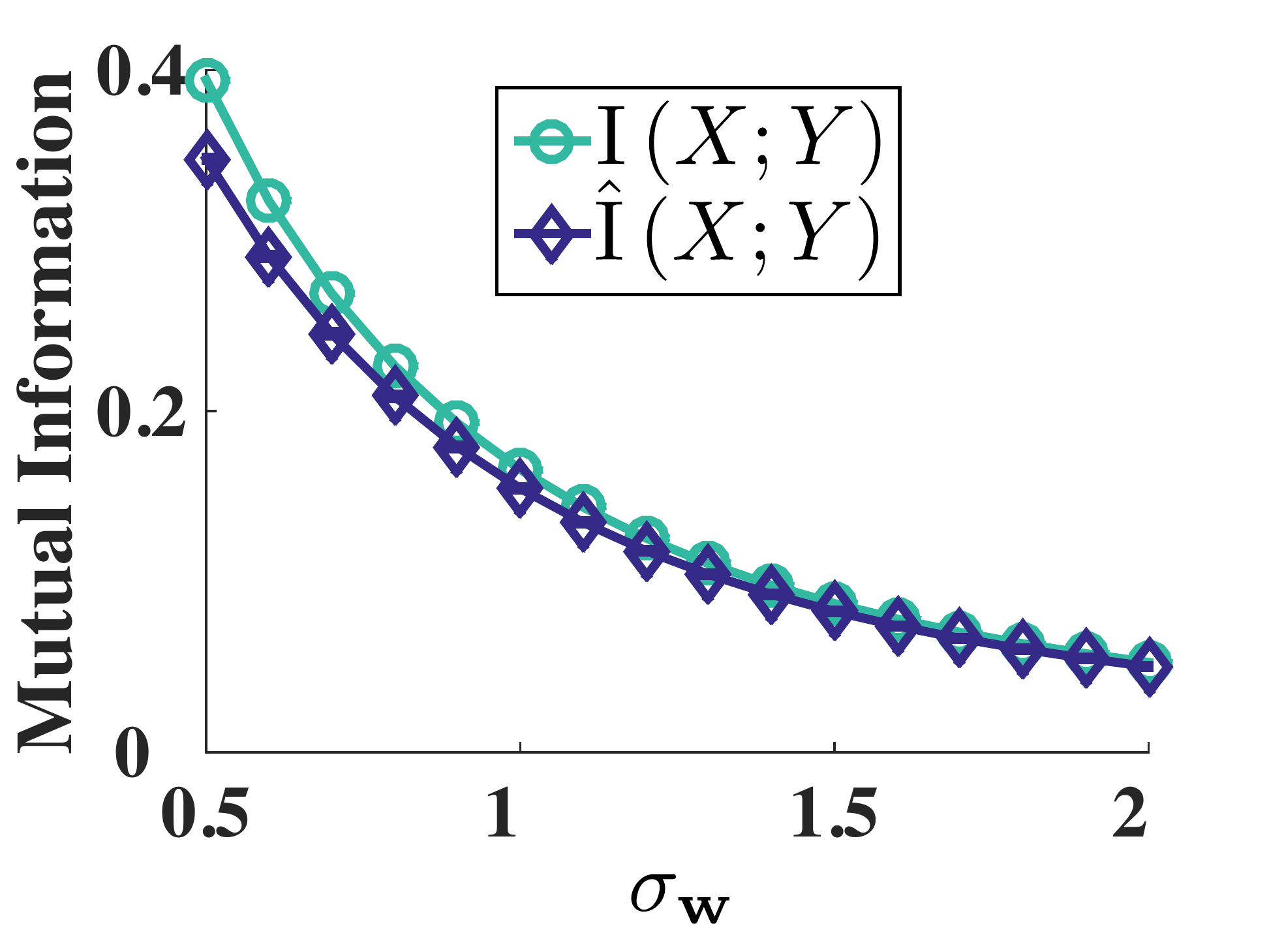} }
\caption{True value of MI between $X$ and $Y$, $\I\left(X;Y\right)$, and the estimate from the proposed algorithm, $\hat{\I}\left(X;Y\right)$, when they are related by (a) an FIR two-tap lowpass filter for different values of $\beta$ (b) a bandpass filter for different values of noise standard deviation, $\sigma_w$. 
}\label{Fig:linear_model}
\squeezeup
\end{figure}

The filter unit-impulse response is $h = \left[\beta, 1-\beta\right]$, for $\beta \in \left[0, 1\right]$. We generated samples of random processes $X$ and $Y$ for different values of $\beta \in \left[0, 1\right]$ with $\sigma_x = \sigma_w = 1$. 
For each value of $\beta \in \left[0, 1\right]$, we calculated the true value of mutual information between $X$ and $Y$, $\I\left(X;Y\right)$, by using the analytical expression\footnote{Note that for this particular model, mutual information is equal to the directed information from $X$ to $Y$ and the analytical expression is given in equation (18) in \cite{malladi2016}.} derived in \cite{malladi2016}. 
The true MI value, $\I\left(X;Y\right)$  and our MI estimate, $\hat{\I}\left(X;Y\right)$, obtained from Algorithm~\ref{mi_est_algo} are plotted for different values of $\beta$ in Fig.~\ref{Fig:linear_model}a. It is seen that the proposed estimator correctly estimates the true value of MI, without the knowledge of the underlying model.

\subsubsection{Bandpass Filter}  \label{sec:bandpass}
We now consider a $33$-tap bandpass filter with passband in the normalized frequency range $\left[0.15, 0.35\right]$. We then generated samples of $X$ and $Y$ from this model for different value of noise power, $\sigma_w \in \left[0.5, 2\right]$ and $\sigma_x = 1$. Note that our estimator does not assume any parametric model for the underlying data and is purely a data-driven estimator. The true value of MI, $\I\left(X;Y\right)$, numerically computed using the power spectral density (chapter 10 in \cite{cover2012}), and the MI estimate, $\hat{\I}\left(X;Y\right)$, obtained from Algorithm~\ref{mi_est_algo} are plotted in Fig.~\ref{Fig:linear_model}b. Again, the proposed MI estimator correctly estimates the MI between these two processes.

\subsection{Nonlinear Models} 
Consider a square nonlinearity wherein the data are generated from 
\begin{align}
y[n] = x[n]^2 + w[n], 
\end{align} 
where $W$ is a white Gaussian noise with standard deviation $\sigma_w$. Computing the true value of mutual information between $X$ and $Y$ numerically is nontrivial. We therefore, use the proposed algorithm to estimate $\hat{\I}\left(X;Y\right)$ for different values of $\sigma_w \in \left[0, 10\right]$, with $N_f=32, N_s=10^4$ and averaged using $10$ different random number generator seeds, to determine if the mutual information estimate is decreasing with increasing $\sigma_w$ as expected. We consider two different models for $X$ such that the samples of $X$ are dependent across time.

\subsubsection{Random Cosine with Squared Nonlinearity} \label{sec:cosine_squared}
\begin{figure}[!t]
\centering
\subfloat[$\widehat{\MI}_{XY}\left( \lambda_i, \lambda_j\right)$]{
\includegraphics[width=0.45\columnwidth]{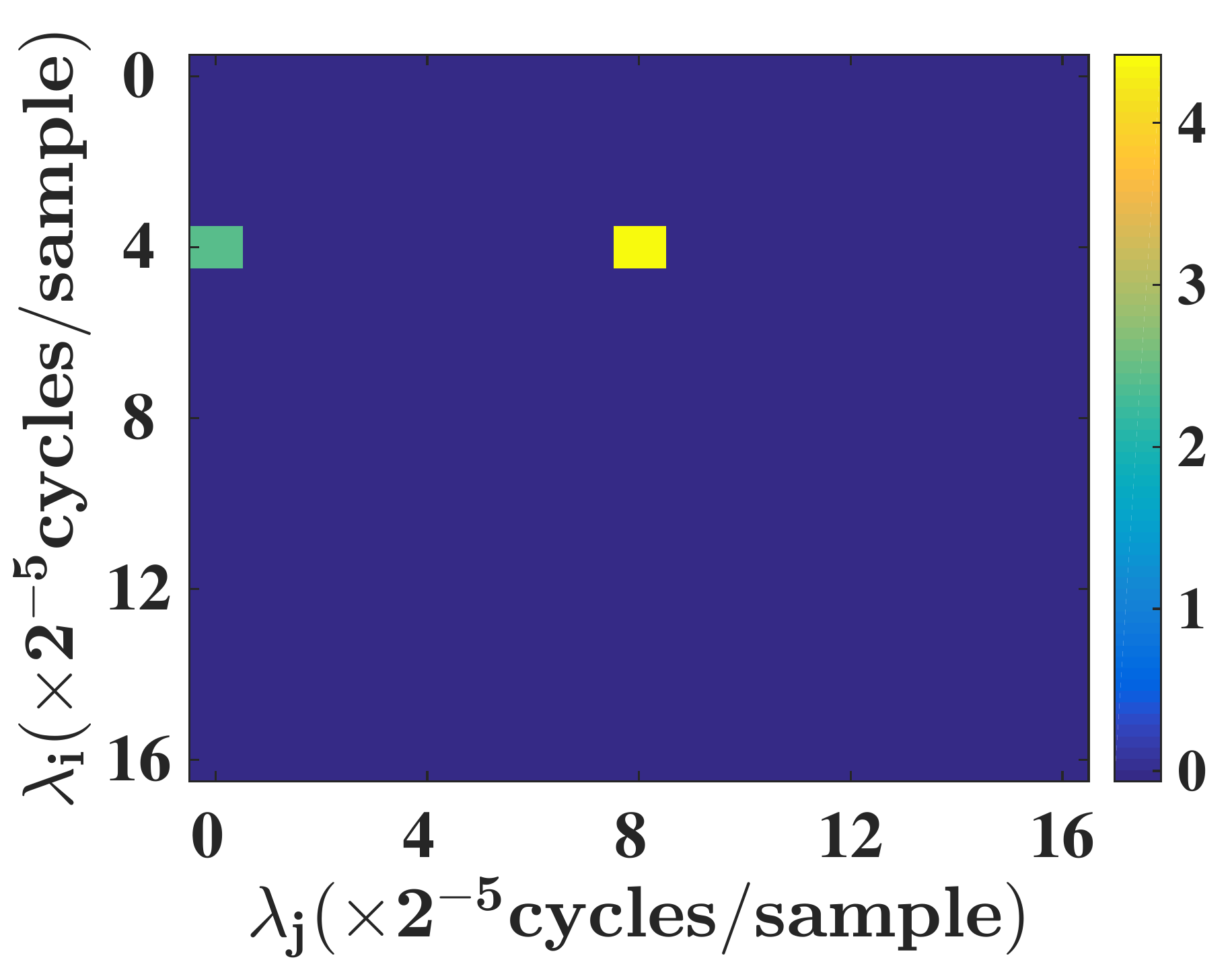}}
\hfill \vrule \hfill
\subfloat[MI between $X$ and $Y$]{
\includegraphics[width=0.45\columnwidth]{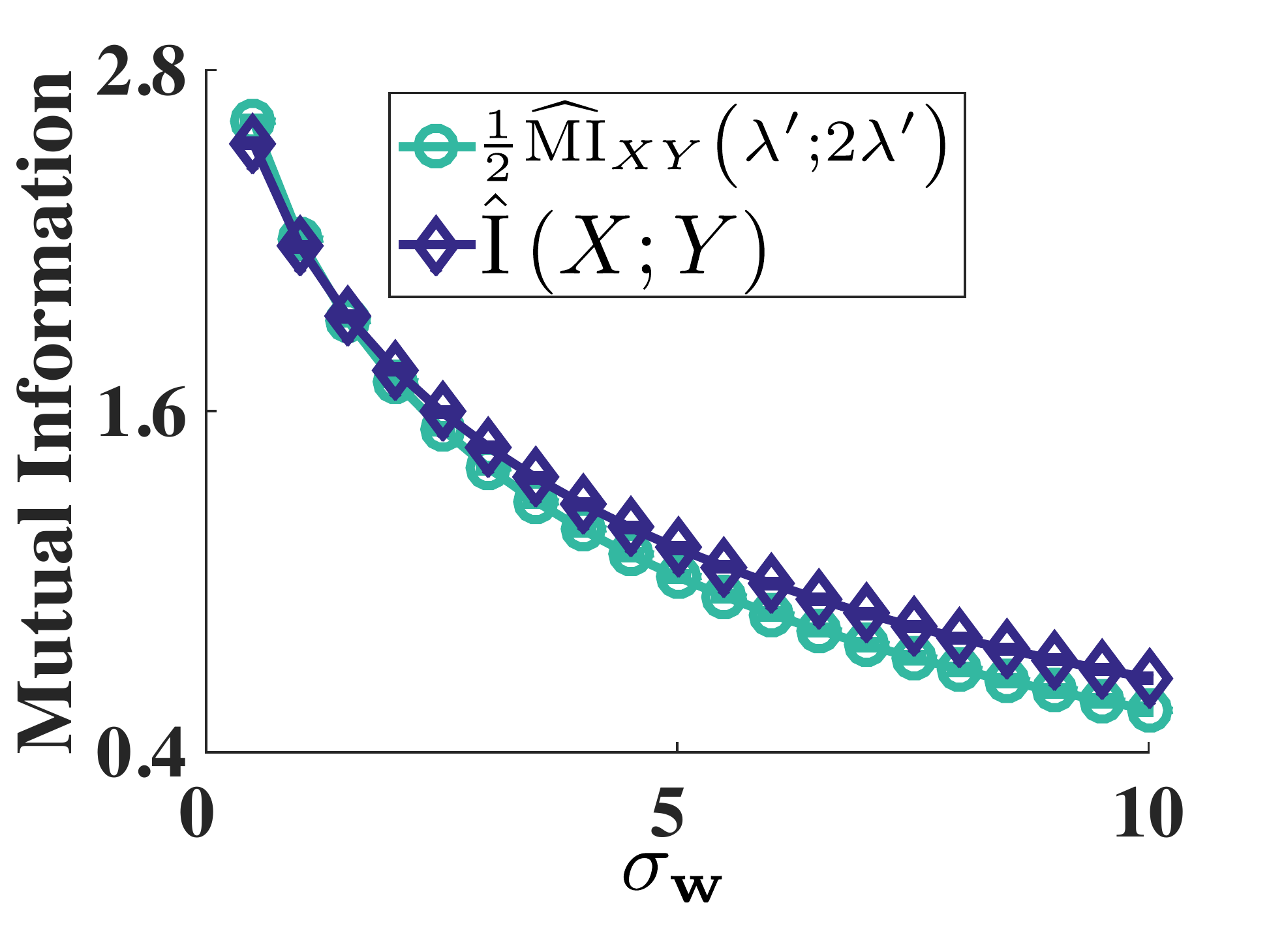} }
\caption{(a) MI in frequency estimates between random processes $X$ and $Y$ related by the single cosine data-generation model. It is clear that MI in frequency estimator correctly identifies the pairwise frequency dependencies. (b) MI in frequency between $X$ at $\lambda^{\prime}$ and $Y$ at $2\lambda^{\prime}$, $\widehat{\MI}_{XY}\left( \lambda^{\prime}, 2\lambda^{\prime}\right)$, obtained from \eqref{eq_knn_est} along with the MI estimate between $X$ and $Y$, $\hat{\I}\left(X;Y\right)$, obtained from Algorithm~\ref{mi_est_algo} for various values of the noise standard deviation, $\sigma_w$. 
}\label{Fig:cosine_square}
\squeezeup
\end{figure}

The samples of $X$ are generated from a random cosine wave, 
\begin{align}
x[n] = A\cos\left(2\pi\lambda^{\prime}n +\theta \right),
\end{align} 
where $A$ is a Rayleigh random variable with parameter $1$, $\theta$ is a uniform random variable between $0$ and $2\pi$ and $\lambda^{\prime} = \frac{4}{32}$. It is very clear that the common information between these two processes will be present between $\lambda^{\prime}$ component of $X$ and the $\left\{0, 2\lambda^{\prime}\right\}$ components of $Y$. This cross-frequency dependence is confirmed by Fig.~\ref{Fig:cosine_square}a, which plots the estimates of pairwise mutual information in frequency between $X$ and $Y$ generated with $\sigma_w=1$ and obtained from \eqref{eq_knn_est}: we observe that significant  dependencies occur only at $\left(\lambda^{\prime}, 0 \right)$ and $\left(\lambda^{\prime}, 2\lambda^{\prime} \right)$ frequency pairs. As a result, $P=1, Q=2$.  The MI estimate from the proposed algorithm, $\hat{\I}\left(X;Y\right) =  \frac{1}{2}\hat{\I}\left(d\widetilde{X}(\lambda^{\prime}); \big\{d\widetilde{Y}(0), d\widetilde{Y}(2\lambda^{\prime})\big\} \right)$ is plotted in Fig.~\ref{Fig:cosine_square}b. The MI estimate decreases with increasing $\sigma_w$ as expected. In addition, we note for this model that the DC component of $Y$ does not contain any extra information about $X$, given the $2\lambda^{\prime}$ component of $Y$. Therefore, we expect  $\frac{1}{2}\hat{\I}\left(d\widetilde{X}(\lambda^{\prime}); \big\{d\widetilde{Y}(0), d\widetilde{Y}(2\lambda^{\prime})\big\} \right) = \frac{1}{2} \widehat{\MI}_{XY}\left(\lambda^{\prime}; 2\lambda^{\prime} \right)$, a result verified in Fig.~\ref{Fig:cosine_square}b, since the two curves are very close. 

\subsubsection{Two Random Cosines with Squared Nonlinearity}\label{sec:two_cosine_squared}
\begin{figure}[!t]
\centering
\subfloat[$\widehat{\MI}_{XY}\left( \lambda_i, \lambda_j\right)$]{
\includegraphics[width=0.45\columnwidth]{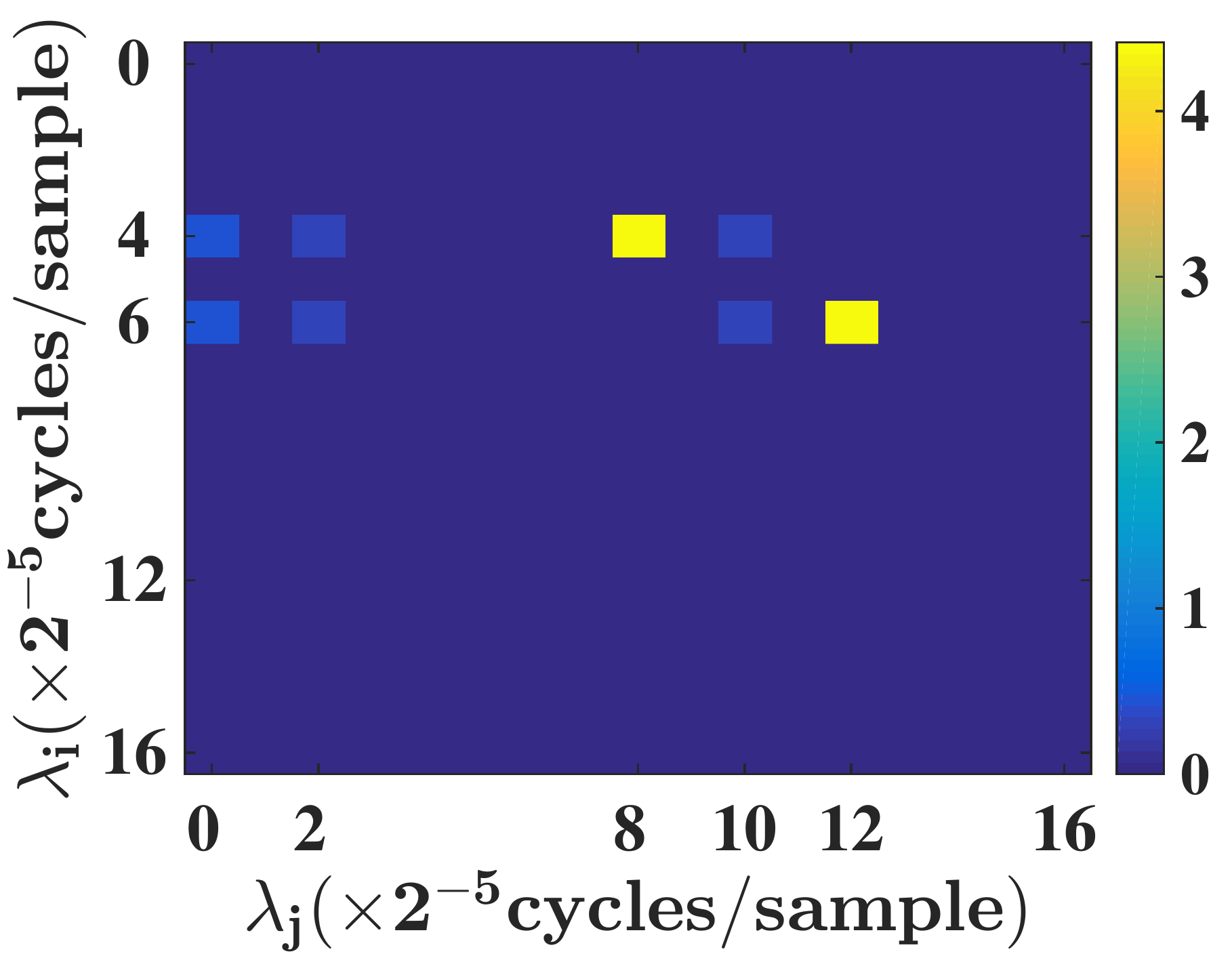}}
\hfill \vrule \hfill
\subfloat[MI between $X$ and $Y$]{
\includegraphics[width=0.45\columnwidth]{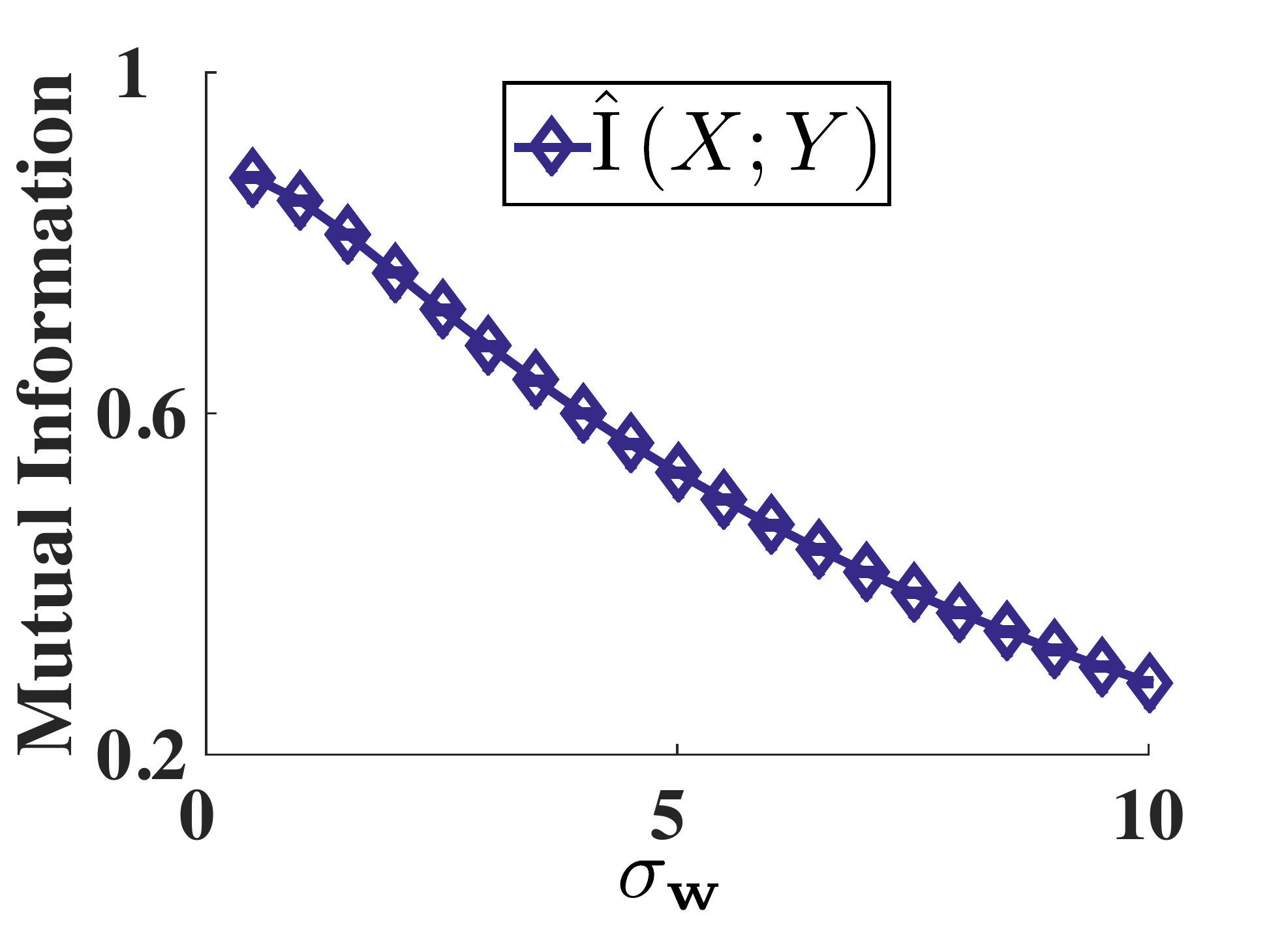} }
\caption{(a) MI in frequency estimates between random processes $X$ and $Y$ related by the two cosine data-generation model. It is clear that MI in frequency estimator correctly identifies the pairwise frequency dependencies between $X$ and $Y$. (b) $\hat{\I}\left(X;Y\right)$, the MI estimate between $X$ and $Y$ obtained from Algorithm~\ref{mi_est_algo} for various values of the noise standard deviation, $\sigma_w$. 
}\label{Fig:two_cosine_square}
\squeezeup
\end{figure}

The samples of random process $X$ are generated according to 
\begin{align}
x[n] = A_1\cos\left(2\pi\lambda^{\prime}n +\theta_1 \right) + A_2\cos\left(2\pi\lambda^{\prime\prime}n +\theta_2 \right),
\end{align} 
where $A_1, A_2$ are independent Rayleigh random variables with parameter $1$, $\theta_1, \theta_2$ are independent uniformly distributed random variables between $0$ and $2\pi$, and $\lambda^{\prime} = \frac{4}{32}, \lambda^{\prime\prime}=\frac{6}{32}$. After some basic algebra, it is easy to see that these pairwise frequency dependencies between $X$ and $Y$ occur at  $\left(\lambda^{\prime}, 0 \right)$, $\left(\lambda^{\prime}, \lambda^{\prime\prime} - \lambda^{\prime} \right)$, 
$\left(\lambda^{\prime}, 2\lambda^{\prime} \right)$,  $\left(\lambda^{\prime}, \lambda^{\prime\prime} + \lambda^{\prime} \right)$, $\left(\lambda^{\prime\prime}, 0 \right)$, $\left(\lambda^{\prime\prime}, \lambda^{\prime\prime} - \lambda^{\prime} \right)$, $\left(\lambda^{\prime\prime}, \lambda^{\prime\prime} + \lambda^{\prime} \right)$ and $\left(\lambda^{\prime\prime}, 2\lambda^{\prime\prime} \right)$.  Fig.~\ref{Fig:two_cosine_square}a plots the estimates of pairwise MI in frequency between $X$ and $Y$ generated with $\sigma_w=1$ and obtained from the data-driven algorithm described in section~\ref{sec:ksg_est}. The algorithm correctly identifies all the dependent frequency pairs and $P=2, Q=5$. We then apply the algorithm described in section~\ref{sec:mi_est} and plot the estimates the MI for different values of noise standard deviation $\sigma_w$ in Fig.~\ref{Fig:two_cosine_square}b. Again, the MI decreases with increasing noise power, as expected. These four different models demonstrate the performance and accuracy of the proposed data-driven MI  estimator.

\section{Conclusions}
In this paper, we developed a data-driven estimator for mutual information between dependent data. The proposed MI estimator converges to the true value for Gaussian data and performs well on data from nonlinear models. The main novelty of the proposed algorithm lies in utilizing frequency domain to estimate a time-domain metric and defining a measure of information in frequency, referred to as mutual information in frequency, that detects and quantifies statistical dependence. Going forward, the performance of the proposed estimator needs to analyzed for specific families of nonlinear relationships in data. In addition, we also successfully applied the mutual information in frequency metric to electrocorticographic recordings from human brain to infer the cross-frequency coupling mechanisms underlying epileptic activity.


\bibliographystyle{IEEEtran}
\bibliography{refs}

\begin{thebibliography}{10}
\providecommand{\url}[1]{#1}
\csname url@samestyle\endcsname
\providecommand{\newblock}{\relax}
\providecommand{\bibinfo}[2]{#2}
\providecommand{\BIBentrySTDinterwordspacing}{\spaceskip=0pt\relax}
\providecommand{\BIBentryALTinterwordstretchfactor}{4}
\providecommand{\BIBentryALTinterwordspacing}{\spaceskip=\fontdimen2\font plus
\BIBentryALTinterwordstretchfactor\fontdimen3\font minus
  \fontdimen4\font\relax}
\providecommand{\BIBforeignlanguage}[2]{{%
\expandafter\ifx\csname l@#1\endcsname\relax
\typeout{** WARNING: IEEEtran.bst: No hyphenation pattern has been}%
\typeout{** loaded for the language `#1'. Using the pattern for}%
\typeout{** the default language instead.}%
\else
\language=\csname l@#1\endcsname
\fi
#2}}
\providecommand{\BIBdecl}{\relax}
\BIBdecl

\bibitem{friston1994}
K.~J. Friston, ``Functional and effective connectivity in neuroimaging: a
  synthesis,'' \emph{Human brain mapping}, vol.~2, no. 1-2, pp. 56--78, 1994.

\bibitem{shannon1948}
C.~E. Shannon, ``A mathematical theory of communication,'' 1948.

\bibitem{cover2012}
T.~M. Cover and J.~A. Thomas, \emph{Elements of information theory}.\hskip 1em
  plus 0.5em minus 0.4em\relax John Wiley \& Sons, 2012.

\bibitem{wang2009}
Q.~Wang, S.~R. Kulkarni, and S.~Verd{\'u}, ``Universal estimation of
  information measures for analog sources,'' \emph{Foundations and Trends in
  Communications and Information Theory}, vol.~5, no.~3, pp. 265--353, 2009.

\bibitem{pinsker1960}
M.~S. Pinsker, ``Information and information stability of random variables and
  processes,'' 1960.

\bibitem{larson1979}
H.~J. Larson and B.~O. Shubert, \emph{Probabilistic models in engineering
  sciences}.\hskip 1em plus 0.5em minus 0.4em\relax Wiley, 1979, vol.~2.

\bibitem{cramer2013}
H.~Cram{\'e}r and M.~R. Leadbetter, \emph{Stationary and related stochastic
  processes: Sample function properties and their applications}.\hskip 1em plus
  0.5em minus 0.4em\relax Courier Corporation, 2013.

\bibitem{malladi2017}
R.~Malladi, D.~Johnson, G.~Kalamangalam, N.~Tandon, and B.~Aazhang, ``Measuring
  cross-frequency coupling using mutual information and its application to
  epilepsy,'' \emph{to be submitted to IEEE Transactions on Signal Processing},
  2017.

\bibitem{aru2015}
J.~Aru, J.~Aru, V.~Priesemann, M.~Wibral, L.~Lana, G.~Pipa, W.~Singer, and
  R.~Vicente, ``Untangling cross-frequency coupling in neuroscience,''
  \emph{Current opinion in neurobiology}, vol.~31, pp. 51--61, 2015.

\bibitem{brillinger2001}
D.~R. Brillinger, \emph{Time series: data analysis and theory}.\hskip 1em plus
  0.5em minus 0.4em\relax Siam, 2001, vol.~36.

\bibitem{scott2015}
D.~W. Scott, \emph{Multivariate density estimation: theory, practice, and
  visualization}.\hskip 1em plus 0.5em minus 0.4em\relax John Wiley \& Sons,
  2015.

\bibitem{kraskov2004}
A.~Kraskov, H.~St{\"o}gbauer, and P.~Grassberger, ``Estimating mutual
  information,'' \emph{Physical review E}, vol.~69, no.~6, p. 066138, 2004.

\bibitem{malladi2016}
R.~Malladi, G.~Kalamangalam, N.~Tandon, and B.~Aazhang, ``Identifying seizure
  onset zone from the causal connectivity inferred using directed
  information,'' \emph{IEEE Journal of Selected Topics in Signal Processing},
  vol.~10, no.~7, pp. 1267--1283, Oct 2016.

\end{thebibliography}

\end{document}